\documentclass[sigconf]{acmart}

\usepackage{algorithm}
\usepackage{algpseudocode}
\usepackage{graphicx}
\usepackage{subcaption}
\usepackage{threeparttable}
\graphicspath{ {./graphics/} }
\usepackage{comment}
\usepackage{etoolbox}
\usepackage{xcolor}
\usepackage{cleveref}
\usepackage{enumitem}
\usepackage[normalem]{ulem}

\AtBeginDocument{%
  }

\setcopyright{acmcopyright}\acmConference[DAC '24]{Proceedings of the 61st ACM/IEEE Design Automation Conference (DAC)}{June 23--27, 2024}{San Francisco, CA, USA}
\acmBooktitle{Proceedings of the 61st ACM/IEEE Design Automation Conference (DAC) (DAC '24), June 23--27, 2024, San Francisco, CA, USA}

\setcopyright{none}
\renewcommand\footnotetextcopyrightpermission[1]{}
\newcommand{\jz}[1]{\ifbool{inccomment}{{\color{red}#1}}{}}
\newcommand{\jk}[1]{\ifbool{inccomment}{{\color{teal}#1}}{}}
\newcommand{\qz}[1]{\ifbool{inccomment}{{\color{blue}#1}}{}}
\newcommand{\ECC}{ECC}

\begin{document}

\title{ModSRAM: Algorithm-Hardware Co-Design for Large Number Modular Multiplication in SRAM}

\author{
    Jonathan Ku\textsuperscript{1},
    Junyao Zhang\textsuperscript{1},
    Haoxuan Shan\textsuperscript{1},
    Saichand Samudrala\textsuperscript{2},
    Jiawen Wu\textsuperscript{2},\\
    Qilin Zheng\textsuperscript{1},
    Ziru Li\textsuperscript{1},
    JV Rajendran\textsuperscript{2},
    Yiran Chen\textsuperscript{1}\\ 
}
\affiliation{%
\textsuperscript{1}Duke University, \textsuperscript{2}Texas A\&M University\\
\country{}
}
\email{jonathan.ku@duke.edu}

\renewcommand{\shortauthors}{
}

\begin{abstract}
Elliptic curve cryptography (ECC) is widely used in security applications such as public key cryptography (PKC) and zero-knowledge proofs (ZKP). ECC is composed of modular arithmetic, where modular multiplication takes most of the processing time. Computational complexity and memory constraints of ECC limit the performance. Therefore, hardware acceleration on ECC is an active field of research. Processing-in-memory (PIM) is a promising approach to tackle this problem. In this work, we design ModSRAM, the first 8T SRAM PIM architecture to compute large-number modular multiplication efficiently. In addition, we propose R4CSA-LUT, a new algorithm that reduces the cycles for an interleaved algorithm and eliminates carry propagation for addition based on look-up tables (LUT). ModSRAM is co-designed with R4CSA-LUT to support modular multiplication and data reuse in memory with 52\% cycle reduction compared to prior works with only 32\% area overhead. 
\end{abstract}


\maketitle

\begin{figure}[t] 
  \centering
  \includegraphics[width=\linewidth]{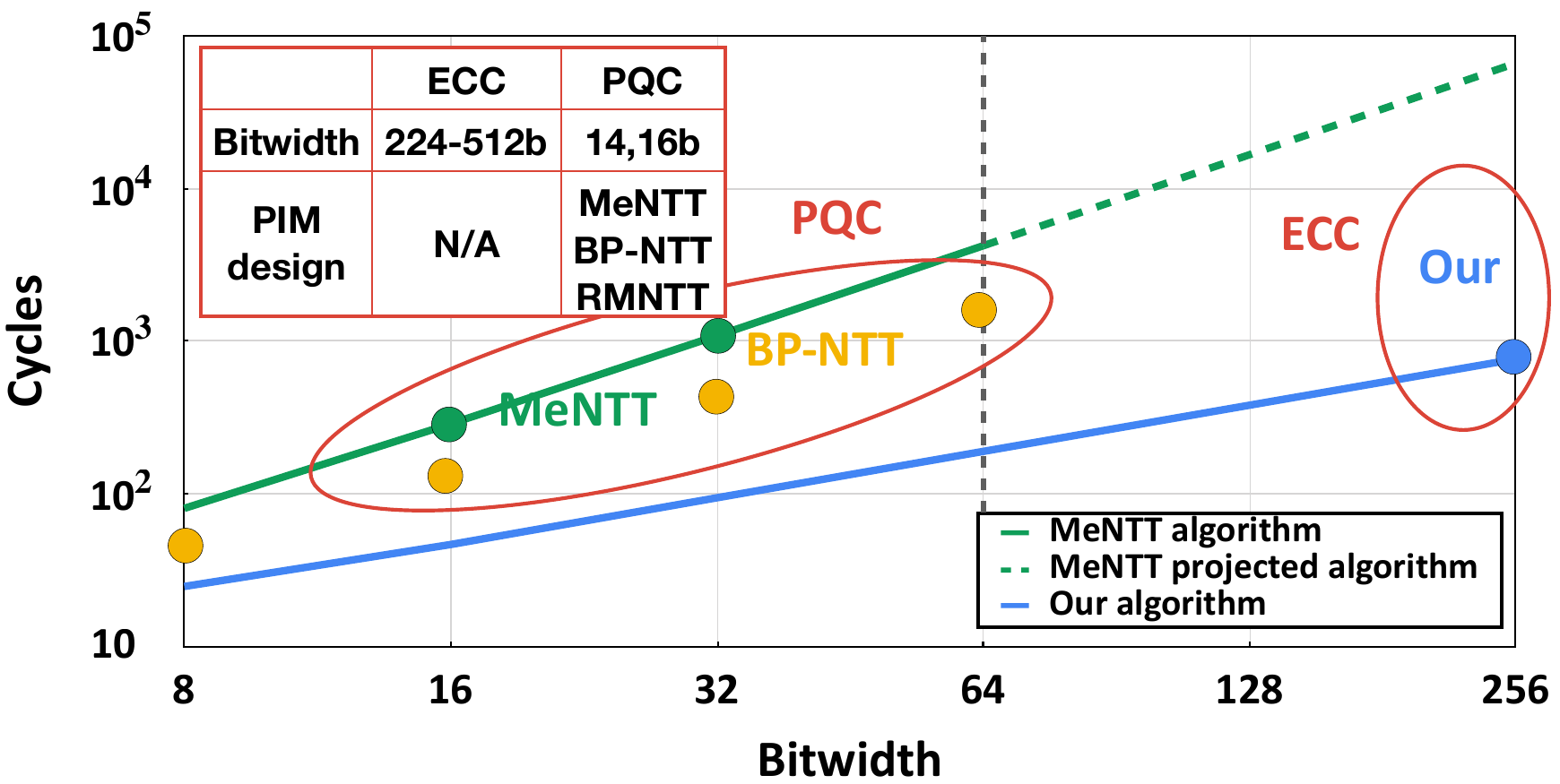}
  \vspace{-20pt}
  \caption{Algorithm complexity and performance comparison with previous work.}
  \label{fig:teaser}
  \Description{A block diagram to show logic-SA.}
  \vspace{-15pt}
\end{figure}
\vspace{-7pt}
\section{Introduction} \label{sec:intro}
Security has become increasingly important in recent years as people care more about privacy and the protection of personal data on the Internet. Public key cryptography (PKC) is commonly used for various applications, such as digital signature and encryption, to name a few. Elliptic curve cryptography (\ECC{}) \cite{ecc} is one of the popular algorithms. It has the benefit of fewer bitwidth for private keys compared to RSA \cite{rsa} with the same security level. Another application that is based on ECC is zero-knowledge proof (ZKP) \cite{zkp}, which is an emerging cryptographic protocol that can prove to the verifier that one statement is true without sharing any secret information other than the statement itself.  

However, ECC needs to perform modular multiplications for operands with a large bitwidth~(at least 224 bits \cite{nist}), and a large number of intermediate results will be generated during the computation process. Thus, deploying the ECC algorithm on the hardware efficiently is a challenging issue due to its high memory bandwidth requirement and high computational complexity.
For example, \cite{pipezk} mentioned it requires 2.98 TB/s bandwidth in 100 MHz to compute a ZKP scheme, which is impractical for the current systems. To mitigate this problem, processing-in-memory (PIM) is an emerging field of research that aims to minimize the gap between computing and memory units. 
Previous works \cite{recryptor,mentt,bpntt,rmntt,cryptopim,xpoly,psram} have demonstrated possible solutions for both SRAM and ReRAM PIM in cryptographic applications ranging from advanced encryption standard (AES), homomorphic encryption (HE) to post-quantum cryptography (PQC). However, none of them target ECC as the computation requires large bitwidth. As shown in \Cref{fig:teaser}, the computation cycles and bitwidth of ECC are higher than PQC. The interface circuit sizes or processing cycles scale up exponentially for large-number modular multiplication. As a result, the existing design methodology for PIM is unsuitable for performing an efficient computation on ECC.

To alleviate the computational complexity problem, in this work, we propose an algorithm-hardware co-design methodology customized for PIM-based architecture. Inspired by previous works \cite{radix4,carrysave}, our proposed algorithm uses a radix-4 encoder and carry save addition features to reduce the computational complexity of the large modular multiplication. In addition, we further customized an SRAM-based PIM architecture to efficiently support the algorithm. Bitwise logic in-memory circuit and simple near-memory circuit features in our proposed SRAM-based PIM architecture provide a significant hardware efficiency improvement due to greater throughput and short critical path.

Overall, our work has the following contributions: 
\setlist{nolistsep}
\begin{itemize}[left=0pt]
\item{We propose R4CSA-LUT, a novel algorithm based on interleaved modular multiplication co-designed with ModSRAM. The latency is greatly improved by using a radix-4 encoder to reduce iterative cycles and employing carry-save addition to eliminate intermediate carry propagation.}

\item{We design ModSRAM, our cryptographic accelerator. It is an 8T SRAM PIM architecture that is co-designed with R4CSA-LUT. ModSRAM utilizes bitwise logic operations to efficiently compute carry save addition in SRAM with simple in/near-memory circuits. Our accelerator is the first to realize large-number modular multiplication in SRAM.}

\item{ModSRAM is implemented and verified through simulation in TSMC 65nm PDK. We have our result through circuit-level simulation and layout, which achieves 52\% fewer cycles with only 32\% area overhead under large bitwidth compared to prior works.}
\end{itemize}


\newpage
\begin{algorithm}[t] \small
  \caption{Interleaved Modular Multiplication} \label{alg:interleaved}
  \footnotesize
  \begin{algorithmic}[1]
    \Require $\text{n-bit } A=(a_{n-1},...,a_0), B, p; 0 \leq A, B \leq p $
    \Ensure $C = A \times B \mod p$
    \State $C \gets 0$
    \For{$a_i \text{ from } a_{n-1} \text{ to } a_0$}
      \State $C \gets 2 \times C$
      \If{$C > p$}
        \State $C \gets C - p$
      \EndIf
      \State $C \gets C + a_i \times B$
      \If{$C > p$}
        \State $C \gets C - p$
      \EndIf    
    \EndFor
  \end{algorithmic}
\end{algorithm}
\vspace{-10pt}

\begin{algorithm}[t] \small
  \caption{Radix-4 Modular Multiplication} \label{alg:radix4} 
  \footnotesize
  \begin{algorithmic}[1]
    \Require $\text{n-bit } A=(a_{n-1},...,a_0), B, p; 0 \leq A, B \leq p$ 
    \Statex $\text{precomputed radix-4 encoding \& overflow LUT}$
    \Ensure $C = A \times B \mod p$
    \State $C \gets 0$
    \For{$i \text{ from } \lceil{\frac{n}{2}-1}\rceil \text{ to } 0$}
      \State $C \gets 4 \times C$
      \If{$C > p$}
        \State $C \gets LUT(C)$
      \EndIf 
      \State $E = ENC(a_{2i+1},a_{2i},a_{2i-1})$
      \State $C \gets C + E \times p$   
      \If{$C > p$}
        \State $C \gets C - p$
      \EndIf 
    \EndFor
  \end{algorithmic}
\end{algorithm}
\vspace{-20pt}

\section{Background and Related Work} 
\label{sec:background}
This section provides the necessary background useful in understanding R4CSA-LUT and previous works on logic PIM \footnote{Logic PIM here is categorized for PIM computing bitwise logic operations, which is in opposition to ML PIM.}. One of the applications for logic PIM is cryptography.  
Even though the target applications from previous works are different than ours, we provide them for completeness.

\vspace{-8pt}
\subsection{Modular Multiplication Algorithms} \label{subsec:modmul}
In modular reduction while doing multiplication, interleaved algorithm \cite{interleaved} shown in \Cref{alg:interleaved} is the fundamental algorithm. It is based on the traditional shift-and-add fashion to do multiplication with a reduction step in every iteration. The total iterations scale with bitwidth, which can be a serious issue in large numbers. Booth-encoded multipliers \cite{booth} are used in modern computers to accelerate multiplication. Instead of iterating through each bit, booth-encoded radix-4 multipliers process three bits at a time with one bit overlapping, which is equivalent to processing two bits in every iteration. Thus, the total iterations are cut in half with the use of an extra encoder. The encoder follows the logic from \Cref{tab:radix4_enc}. Radix-8 multipliers are very similar. Four bits are processed with one bit overlapping. As a result, the total iterations are cut down by one-third. The idea of a booth-encoded multiplier can be integrated with the interleaved algorithm as shown in \Cref{alg:radix4}. Hardware implementation results are shown in \cite{radix4} with significant iteration reduction.  

The work \cite{carrysave} proposed a carry-save addition-based interleaved algorithm to improve performance. For every loop in \Cref{alg:interleaved}, there is a shift, two comparisons followed by subtractions, and a full addition. Shift induces an extra reduction step (comparison then subtraction) since the result is doubled and full addition induces carry propagation, thus increasing hardware resources and latency. 

\begin{table}[t] \small
  \caption{Radix-4 Computation Tables} \label{tab:radix4}
  \footnotesize
  \vspace{-10pt}
  \begin{subtable}{.5\linewidth} 
    \centering
    \caption{Radix-4 Booth Encoder}
    \label{tab:radix4_enc}
    \vspace{-4pt}
    {\begin{tabular}[t]{ccc|c}
      \toprule
      $a_{i+1}$ & $a_i$ & $a_{i-1}$ & ENC\\
      \midrule
      0 & 0 & 0 & 0 \\
      0 & 0 & 1 & +1\\
      0 & 1 & 0 & +1\\
      0 & 1 & 1 & +2\\
      1 & 0 & 0 & -2\\
      1 & 0 & 1 & -1\\
      1 & 1 & 0 & -1\\
      1 & 1 & 1 & 0 \\
      \bottomrule
    \end{tabular}}
  \end{subtable}%
  \begin{subtable}{.5\linewidth}    
    \centering
    \caption{Radix-4 Precomputation LUT}
    \label{tab:LUT-radix4}
    \vspace{-4pt}
    \begin{tabular}[t]{c|c}
      \toprule
      ENC & LUT-radix4\\
      \midrule
      0  & $0$\\
      +1 & $B$\\
      +2 & $2 \times B \mod p$\\
      -2 & $-2 \times B \mod p$\\
      -1 & $-1 \times B \mod p$\\
      \bottomrule
    \end{tabular}
  \end{subtable}
  \vspace{-15pt}
\end{table}

To mitigate this issue, the shift can be considered as adding a new value that is the original value after reduction. The new value can be determined by an extra bit induced by the shift, which we call carry overflow. Since the intermediate results are not our concern, we can adopt carry-save addition to replace full addition. This makes the operation much easier to implement in hardware.

\vspace{-8pt}
\subsection{Cryptographic PIM} \label{subsec:pim}
Recently, there have been many cryptographic PIM accelerators in SRAM \cite{recryptor,mentt,bpntt,psram} and ReRAM \cite{rmntt,cryptopim,xpoly} that tried to compute cryptographic schemes in/near-memory. Among all these works, AES and PQC are the most popular. The basic operations in AES are bitwise logic and shift, which are proposed in many logic PIMs \cite{xsram,bitparallel,psram} already. HE and PQC, on the other hand, is a rising field. HE is an encryption scheme that allows computation directly on ciphertext where plaintext after deciphered, is computed as well. PQC is the field for encryption algorithms that are safe from quantum attacks. No matter the target application, they are all based on polynomial computation, which is usually computed via number theoretical transform (NTT) \cite{ntt}. It is a generalization for discrete Fourier transform (DFT) over a finite field. The basic operation to do so is modular arithmetic. For these applications, the accelerators are designed for small bitwidth, commonly in 14/16-bit. These designs do not scale with bitwidth in applications such as ECC, where the security level recommended to date is at least 224 bits \cite{nist}.

\begin{figure}[b] 
  \vspace{-15pt}  
  \begin{subfigure}[t]{.35\linewidth}
    \centering
    \includegraphics[width=\linewidth]{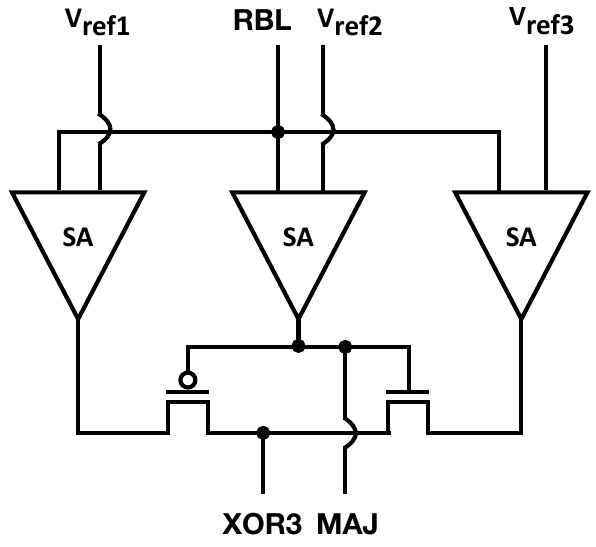}
    \caption{Logic-SA module} \label{fig:logic-sa-mod}
    \Description{A block diagram to show logic-SA.}
  \end{subfigure}%
  \begin{subfigure}[t]{.45\linewidth}
    \centering
    \includegraphics[width=\linewidth]{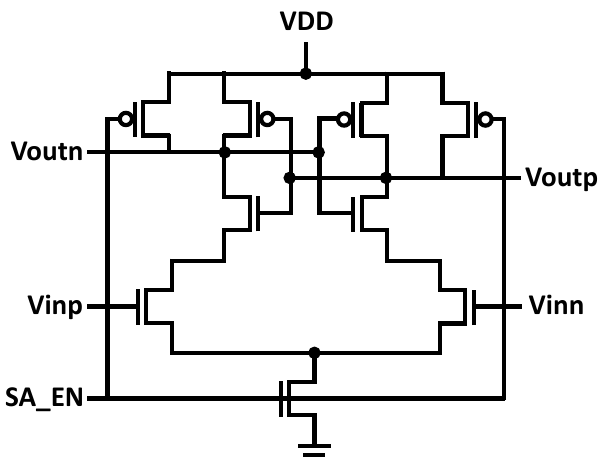}
    \caption{SA} \label{fig:sa}
    \Description{A circuit diagram to show latchtype SA.}
  \end{subfigure}
  \vspace{-10pt}    
  \caption{Logic-SA module for addition proposed in \cite{psram} and latch-type SA structure \cite{sa}.} \label{fig:logic-sa}
\end{figure}

Since the operation in cryptographic PIM can further decompose into bitwise logic and simple logic near-memory, architectures from logic PIM provides the basic design. 2-input logic operations in SRAM are supported in previous works \cite{xsram,bitparallel} and 3-input logic operations in SRAM are first implemented in \cite{psram}. It is the first to realize XOR3 and MAJ (majority) logic functions, which are the sum and carry for addition. The logic-SA module they proposed is illustrated in \Cref{fig:logic-sa}.


\newpage
\begin{algorithm}[H] \small
  \caption{Proposed Modular Multiplication in-SRAM} \label{alg:proposed} 
  \footnotesize
  \begin{algorithmic}[1]
    \Require $\text{n-bit } A=(a_{n-1},...,a_0), B, p; 0 \leq A, B \leq p$ 
    \Statex $\text{LUT-radix4 \& LUT-overflow}$
    \Ensure $C = A \times B \mod p$
    \State $\text{sum} \gets 0$ \Comment{remain n+1 bits}
    \State $\text{carry} \gets 0$ \Comment{remain n+1 bits}
    \For{$i \text{ from } \lceil{\frac{n}{2}-1}\rceil \text{ to } 0$}
      \State $[\text{overflow}_\text{sum}, \text{sum}] \gets \text{sum} << 2$
      \State $[\text{overflow}_\text{carry}, \text{carry}] \gets \text{carry} << 2$
      \State $\text{overflow} \gets \text{overflow}_\text{sum} + \text{overflow}_\text{carry}$ + MSB(LUT-radix4)
      \State $\text{sum} \gets \text{XOR3}(\text{LUT-radix4}(a_{i+1},a_i,a_{i-1}), \text{sum, carry})$
      \State $\text{carry} \gets \text{MAJ}(\text{LUT-radix4}(a_{i+1},a_i,a_{i-1}), \text{sum, carry})$
      \State $\text{carry} \gets \text{carry} << 1$
      \State $\text{sum} \gets \text{XOR3(LUT-overflow(overflow), sum, carry})$
      \State $\text{carry} \gets \text{MAJ(LUT-overflow(overflow), sum, carry})$
      \State $\text{carry} \gets \text{carry} << 1$
    \EndFor
    \State $C \gets \text{sum + carry}$
  \end{algorithmic}
\end{algorithm}
\vspace{-16pt}

\section{R4CSA-LUT Algorithm} \label{sec:algo}
Modular multiplication algorithms can be generalized into two groups as mentioned in \Cref{sec:intro}. Montgomery reduction \cite{montgomery} and Barrett reduction \cite{barrett} are the two most popular methods in reduction after multiplication. Montgomery reduction avoids carry propagation and prevents expensive modular operation by first transforming the operands into Montgomery form. The computations in the Montgomery form are much easier than in its direct form. As a result, the speedup in Montgomery reduction is obvious. Barrett reduction uses another multiplication in place of division for modular reduction. Unfortunately, both of them involve n-bit multiplication, resulting in 2n-bit intermediate results that require more hardware resources to store and compute. In addition, Montgomery reduction requires extra transformation into and out of Montgomery form, which is an unavoidable real modular operation. Barrett reduction induces a 3n-bit intermediate result after the regular multiplication for modular reduction, which takes up even more hardware resources. Both of them reduce the computational latency at the cost of a very complex circuit and memory design in tradeoff. Interleaved modular multiplication \cite{interleaved}, on the other hand, is a potential hardware-friendly solution for reduction while doing multiplication. Numerous algorithms have been proposed based on interleaved algorithm as in \Cref{subsec:modmul}. The proposed algorithm overview and the mapping to our hardware will be discussed.

\vspace{-8pt}
\subsection{Algorithm Overview} \label{subsec:proposed}
In view of the strengths and weaknesses of previous works, we proposed a new algorithm combining the merits of each algorithm called radix-4 carry save addition, a look-up table based interleaved algorithm (R4CSA-LUT). Since the classical interleaved algorithm has long latency due to a large number of iterations, radix-4 modular multiplication in \Cref{alg:radix4} is adopted in R4CSA-LUT to cut iterations in half with only an extra booth encoder as in \Cref{tab:radix4_enc}. The value added every iteration can be precomputed as in \Cref{tab:LUT-radix4} since there are only five possible values and only three of them need computation. These results can be reused as long as the multiplicand remains the same.  However, \Cref{alg:radix4} still suffers from carry propagation. This issue seriously affects performance when the numbers to be multiplied become larger. Carry save addition can be adopted into the original radix-4 modular multiplier to eliminate long carry propagation latency as previously mentioned. The values for carry overflow can also be precomputed for eight possible cases. They can be reused as long as the modulo number remains the same. R4CSA-LUT is shown in \Cref{alg:proposed}. It achieves half iterations compared to an interleaved algorithm without carry propagation via carry-save addition. It is co-designed with our architecture so that the operations are hardware-friendly and data can be reused through LUT, which will be introduced in \Cref{sec:arch}.

\vspace{-8pt}
\subsection{Mapping to Hardware} \label{subsec:mapping}
The algorithm can be separated into three parts: precomputation, main iteration and computation for the final result. Precomputation can be stored for later use during the main iteration. The LUTs required to store precomputation results are represented in \Cref{tab:LUT-radix4,tab:LUT-overflow}, which are stored in each wordline (WL) in SRAM. The sum and carry overflow can be used to determine the value added for the next cycle. It depends on the most significant four bits of sum, carry, and the most significant bit (MSB) of radix-4 LUT. They can be computed with a rather low cost compared to the whole modular multiplication because their bitwidths are at most n+3 bits. These results can be reused over multiple iterations and multiple calculations, thus reducing memory movement and maximizing data reuse. For the main iteration, carry save addition is the essential operation and bitwise XOR3 and MAJ logic functions represent the sum and carry, respectively. The left shift by two is due to processing two bits in radix-4 modular multiplication. 3-input logic functions are made possible to compute in-memory by logic-SA module \cite{psram} in \Cref{fig:logic-sa}. This provides the fundamental building block to realize R4CSA-LUT in SRAM. The final step is a full addition of the sum and carry in n+1 bits with a reduction step to get the final value. This is inevitable and is best to be computed near-memory. Combining all previous parts, we get our proposed algorithm that can run efficiently on our designed hardware. 

\begin{table}[t] \small
  \caption{Carry Overflow Precomputation LUT} \label{tab:LUT-overflow}
  \scriptsize
  \begin{tabular}{ccc|c}
    \toprule
    $a_{n+3}$ & $a_{n+2}$ & $a_{n+1}$ & LUT-overflow\\
    \midrule
    0 & 0 & 0 & 0\\
    0 & 0 & 1 & $001\overbrace{(0...0)}^{n+1 \rm\ \text{bits}} \mod p$\\
    0 & 1 & 0 & $010\;(0...0)\; \mod p$\\
    0 & 1 & 1 & $011\;(0...0)\; \mod p$\\
    1 & 0 & 0 & $100\;(0...0)\; \mod p$\\
    1 & 0 & 1 & $101\;(0...0)\; \mod p$\\
    1 & 1 & 0 & $110\;(0...0)\; \mod p$\\
    1 & 1 & 1 & $111\;(0...0)\; \mod p$\\
  \bottomrule
  \end{tabular}
\vspace{-10pt}
\end{table}

\begin{figure*}[t] 
  \centering
  \includegraphics[width=\linewidth]{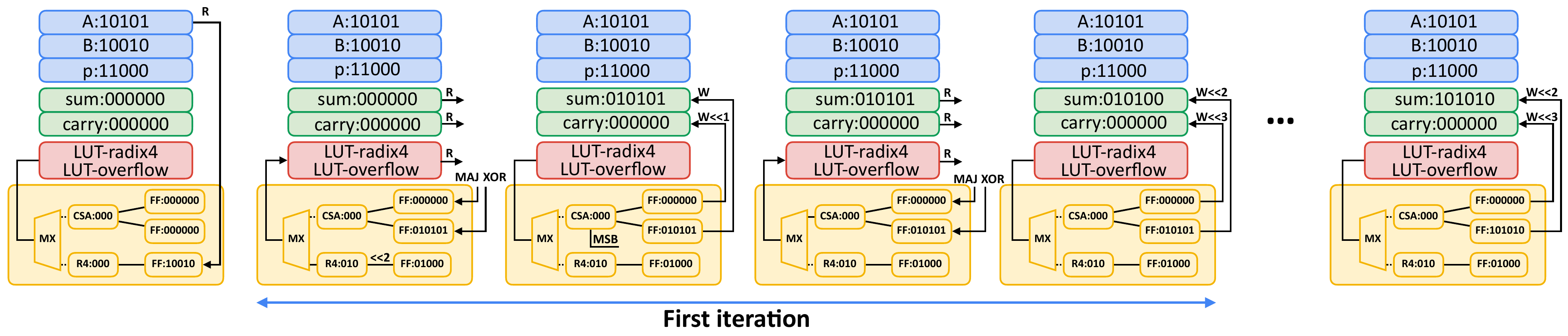}
  \vspace{-20pt}
  \caption{A 5-bit illustration of the first iteration in R4CSA-LUT dataflow with proposed ModSRAM.}
  \label{fig:alg-ill}
  \vspace{-12pt}
\end{figure*}

\begin{figure}[t] 
  \centering
  \includegraphics[width=\linewidth]{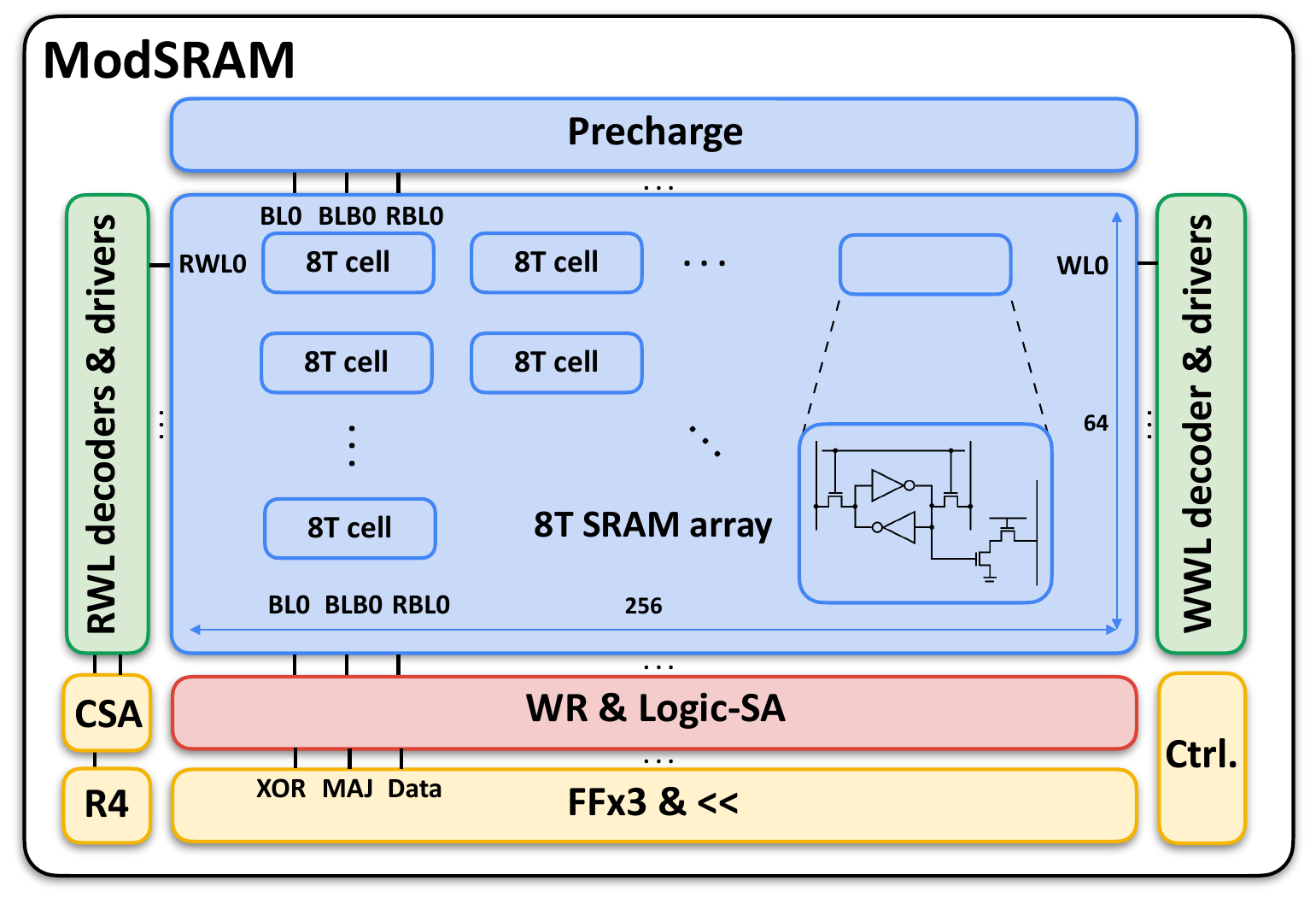}
  \vspace{-22pt}
  \caption{The overall architecture of ModSRAM.}
  \vspace{-15pt}
  \label{fig:arch}
\end{figure}
\begin{figure*}[t] 
  \begin{subfigure}[t]{.2\linewidth}
    \centering
    \includegraphics[width=1\linewidth]{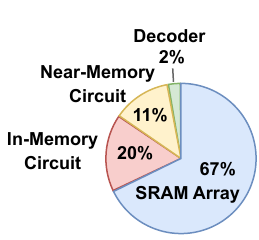}
    \Description{Pie chart for ModSRAM area breakdown.}
  \end{subfigure}
  \begin{subfigure}[t]{.78\linewidth}
    \centering
    \includegraphics[width=\linewidth]{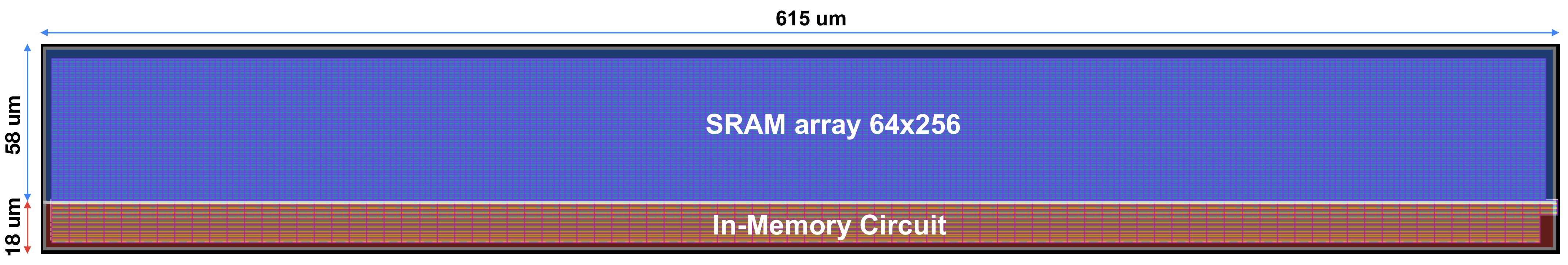}
    \Description{A circuit diagram to show latchtype SA.}
  \end{subfigure}
  \vspace{-5pt}    
  \caption{Area breakdown on ModSRAM and full custom layout for SRAM array and in-memory circuit.} \label{fig:area}
\end{figure*}
\vspace{-7pt}
\section{ModSRAM Architecture} \label{sec:arch}

\subsection{Architecture Overview}
\Cref{fig:arch} illustrates the overall architecture of ModSRAM. It is an SRAM PIM design with custom in/near memory computing circuits to execute the R4CSA-LUT algorithm, which aims to compute modular multiplication in 256 bits efficiently. ModSRAM consists of a 64x256 8T SRAM array with a read port and a write port. The in-memory computing (IMC) circuit is the logic-SA module used to implement XOR3 and MAJ bitwise logic function for carry save addition discussed in detail in \Cref{subsec:imc}. The rest of the peripheral circuits include read wordline (RWL) and write wordline (WWL) decoders as well as near-memory computing (NMC) circuits. They are a radix-4 encoder, combinational logic for overflow, three D flip-flops (DFF) for sum, carry, multiplicand and a controller (Ctrl.), which will be discussed in \Cref{subsec:nmc}. 

\subsection{In-Memory Computing} \label{subsec:imc}
The IMC part includes precharges, SRAM array and a modified sense amplifier (SA) block to enable logic operation. The SRAM cell is standard 8T that supports one read port and one write port. We adopt this design because our algorithm is based on XOR3 and MAJ logic operations, which are three-input logic operations that open three WLs simultaneously. Traditional 6T SRAM suffers from read disturb since read and write share one single port. This issue is even worse when activating two WLs to enable IMC. Since three WLs will be activated simultaneously in our design, read disturbance is no longer negligible. Therefore, a separate read port is necessary to prevent read disturbance while improving read latency. We adopt the logic-SA module shown in \Cref{fig:logic-sa-mod} from \cite{psram}. Three SAs are used for each read bitline (RBL) to differentiate RBL voltage levels for all the 3-input logic functions in this module, with a total of 256 RBLs. SAs used in ModSRAM are conventional voltage-based latch-type sense amplifiers.  

\vspace{-10pt}
\subsection{Near-Memory Computing} \label{subsec:nmc}
Outputs from the IMC circuit are sent to the NMC circuit. They are first stored in FFs, shifted and written back to SRAM for the next iteration. Part of the bits are used to do computation and pass through a MUX to select LUT. To start the iteration, the multiplier is read from SRAM to the near-memory FF. To get the radix-4 encoded computation results in LUT, we take the most significant three bits of the multiplier fetched and encode the following \Cref{tab:radix4}. For every iteration, the multiplier is shifted to the left by two to get the next value for encoding. 

The whole iteration can be partitioned into two sections, which include the first operation for radix-4 LUT and the second operation for overflow LUT. They basically follow the same dataflow, except the data retrieved are in different LUTs, which are different WLs in SRAM. The dataflow for near-memory components is as follows. First, the sum and carry from the previous iteration are shifted to the left by two bits, namely multiplying by four. The overflow bits are stored in a temporary FF for computation in the second section. Next, the encoded result mentioned previously is used to activate WL in radix-4 LUT along with sum and carry. The result from IMC is written back to SRAM with sum first and carry second because during the writeback of sum, carry will be shifted to the left by one bit due to the nature of carry. The overflow bits calculated at the beginning are used to activate WL in overflow LUT along with sum and carry. The result again follows the same datapath. 

After the last iteration, we will get n+1 bits of sum and carry, which requires a full addition and reduction to get our final value. However, since the bitwidth is reduced, this step is rather cheap compared to 2n bits without reduction while doing multiplication. The whole NMC circuit is compact as there are only shifters, three full-bitwidth FFs for the multiplier, sum, carry, and some negligible FFs for overflow. Controller for all SRAM operations such as precharge, activating WLs, 
enabling SA and FSM for near-memory are all realized via Verilog.

\vspace{-8pt}
\subsection{Algorithm Illustration} \label{subsec:ill}
A simplified version of the 5-bit R4CSA-LUT demonstration on ModSRAM is illustrated in \Cref{fig:alg-ill}. For 5-bit modular multiplication, there are three iterations. In \Cref{fig:alg-ill}, only the first iteration is shown. The first step is to read multiplier A into near-memory FF. Then it will be left shifted by two to select the WL in radix-4 LUT. Three WLs are activated at the same time for IMC. The results of IMC are XOR and MAJ, which will be stored in FFs. They are then left-shifted and written back to SRAM. The next step follows the same, except this time overflow LUT is used for IMC. The final results are shown in the end.

\vspace{-8pt}
\section{Evaluation and Discussion} \label{sec:eval}

\subsection{Evaluation Methodology}
We evaluate ModSRAM using TSMC 65nm technology PDK. Full-custom circuits including SRAM array and IMC modules are designed in Cadence Virtuoso. Digital circuits including WL decoders, NMC modules, and a controller are designed in Verilog, and synthesized in Synopsys Design Compiler. Simulations are done in both HSPICE as well as Verilog testbench to get the experimental results. A full-custom layout and synthesis result are included in the analysis to get the design area. The area breakdown and full-custom layout are shown in \Cref{fig:area}.      


\vspace{-8pt}
\subsection{Memory Utilization}
Since we aim for ECC applications, the security level recommended by NIST is at least 224 bits \cite{nist}. Among all the popular elliptic curves (EC), Secp256k1 and BN254 are used for Bitcoin and Zcash, respectively. As a result, we chose 256-bits to be our target. Each WL stores an operand that can be either multiplicand, multiplier, or modulo. Our design is accommodated to fit operands of a point addition operation in EC which are composed of several modular multiplications. During the computation stage, only sum and carry are considered intermediate results that need to be stored in SRAM. Radix-4 and overflow LUTs require a total of 13 WLs, but they can be reused for multiple iterations and for multiple calculations, thus not considered intermediate results. \Cref{fig:data-org} shows the memory utilization comparison for operand storage and intermediate of our work along with existing SRAM PIMs \cite{mentt,bpntt}. LUTs are introduced in our work as shown. 

\vspace{-8pt}
\subsection{Experimental Results}
The number of clock cycles for doing one modular multiplication is recorded in \Cref{tab:cmp}. For 256-bit, it can be done in 767 cycles with the clock frequency given as 420 MHz. R4CSA-LUT algorithm has a complexity of O(n), which scales linearly to bitwidth. The computation result is in the direct form, so no extra conversion cost is needed. The area achieved is small since it only demonstrates the operation of one modular multiplication. The area breakdown in \Cref{fig:area} shows that the memory array occupies two-thirds of the whole design. SAs constitute most of the area in the in-memory circuits with the area of MUX as two transistors negligible. Since our design computes in-memory, the near-memory circuit is compact with very small WL decoders. ModSRAM induces only 32\% area overhead by including near-memory circuits and two SAs since the regular SRAM design includes a WL decoder and an SA already.

\begin{table*} \small
  \begin{threeparttable} 
  \caption{Comparison on modular multiplication in PIM designs.} \label{tab:cmp}
  \vspace{-10pt}
  \begin{tabular}{c||c|c|c|c|c|c}
    \hline
    Reference         &This work&MeNTT \cite{mentt}&BP-NTT \cite{bpntt}&RM-NTT \cite{rmntt}&CryptoPIM \cite{cryptopim}&X-Poly \cite{xpoly}\\
    \hline
    application type  &ECC      &PQC NTT   &PQC NTT        &HE NTT         &PQC NTT               &PQC NTT   \\
    Computation method&direct   &direct    &Montgomery     &Montgomery     &Montgomery/Barrett    &Barrett   \\
    technology        &65 nm    &65 nm     &45nm           &28nm           &45 nm                 &45nm      \\
    Cell type         &8T SRAM  &6T SRAM   &6T SRAM        &ReRAM          &ReRAM                 &ReRAM     \\
    Array size        &64x256   &4x162x256 &4x256x256      &64x4x128x128   &512x512               &16x128x128\\
    Frequency(MHz)    &420      &151       &3.8k           &400            &909                   &400       \\
    Bitwidth          &256      &14/16/32  &2/4/8/16/32/64 &14/16          &16/32                 &16        \\
    Cycles \tnote{*}  &767      &66049     &1465           &-              &-                     &-         \\
    Area $(mm^2)$     &0.053    &0.36      &0.063          &-              &0.152                 &0.27      \\
    \hline
  \end{tabular}
  \begin{tablenotes}
  \scriptsize
  \item[*] Cycles for other works are generated from frequency, latency and number of modular multiplication in NTT scaled to a fair comparison in the same bitwidth (256b).  
  \end{tablenotes}  
  \end{threeparttable}
  \vspace{-10pt}
\end{table*}

\vspace{-8pt}
\subsection{Comparison with existing PIM works}
Even though no PIM works currently implement large bitwidth modular multiplication for ECC applications, some works demonstrated the possibility of PQC NTT \cite{mentt,bpntt,rmntt,cryptopim,xpoly}. The problems in previous works motivated our work. Their number of cycles of a single modular multiplication are scaled to meet our bitwidth and compared. The rest of the experimental data are extracted directly from the works, which are shown in \Cref{tab:cmp}.

\begin{figure}[t] 
  \centering
  \includegraphics[width=\linewidth]{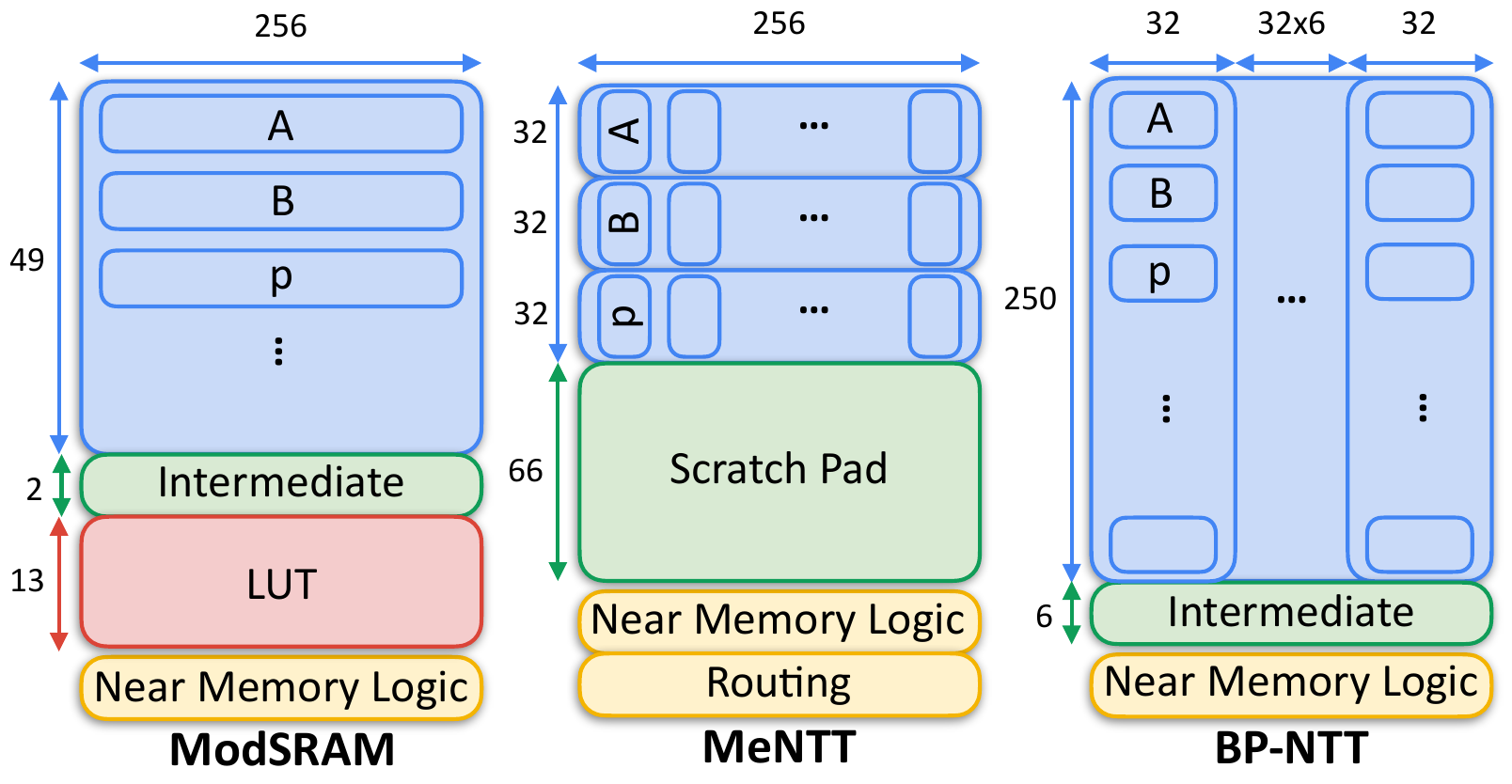}
  \vspace{-20pt}
  \caption{Comparison of data organization for different SRAM PIM designs for modular multiplication.}
  \label{fig:data-org}
  \vspace{-10pt}
  \Description{Three block diagrams for each SRAM PIM design.}
\end{figure}

Regarding SRAM PIM works, \cite{mentt} is one of the first SRAM PIMs in PQC NTT. Their access pattern is bit serial as shown in \Cref{fig:data-org}, meaning that the data is stored across the same BL instead of across WL in order to match with their algorithm. This design faces difficulties when scaling the bitwidth because all the operands are stored in the same BL. Doing the computation in 256 bits requires a total of 1282 rows, which is impractical for an SRAM bank. The corresponding algorithm needs $(n+1)^2$ cycles shown in \Cref{fig:teaser} compared to $3n-1$ cycles in our work. Another work \cite{bpntt} improved the performance by adopting a bit-parallel algorithm. It applies the Montgomery transform to avoid carry propagation in their NTT computation. However, the major issue in this design is the transformation cost. They assumed the precomputation of the Montgomery transform for the operands was readily available before they used the inputs in their PIM. However, when the bitwidth increases, the transformation cost is no longer negligible. 

As for ReRAM PIM works, \cite{cryptopim} introduced PQC NTT with three possible values to choose for modulo. This simplifies the computation yet limits the generality utilized on other applications. \cite{rmntt,xpoly} on the other hand, solved this issue by providing the modulo as an input. They achieved low latency for NTT at the cost of a large design done only in a simulator instead of in circuit-level simulation. The computations are done with modular reduction after multiplication, therefore no cycle results are presented. To accommodate the need for lossless IMC, both designs required a huge area for analog-to-digital converters (ADC) that occupied more than 70\% of the total architecture.

\vspace{-6pt}
\section{Future Work} \label{sec:future}
This work focuses on the design of a modular multiplier. The goal is to reduce latency and area used in large-scale cryptographic applications by utilizing memory and computing components in-memory. The increases in reusability and compactness make it a desired prototype for further research. This paper serves as a pioneer work on realizing large bitwidth modular operation in-memory that was not possible previously. 

\begin{figure}[t] 
  \centering
  \includegraphics[width=.8\linewidth]{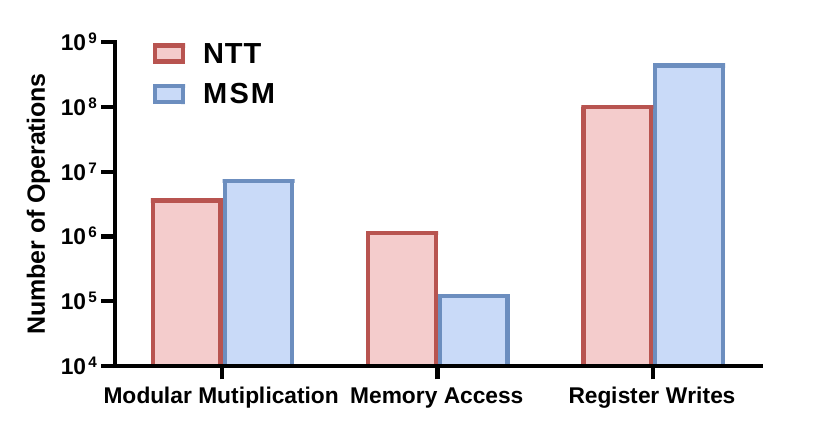}
  \vspace{-15pt}
  \caption{Illustration of the number of operations in ZKP components: NTT~\cite{PNTT}, and MSM ~\cite{pipezk}, when the input vector is of size $2^{15}$ and each input bitwidth is $256$ bits.}
  \label{fig: ntt-msm-eval}
  \Description{Demonstration of the number of operations (256-bit) in ZKP Components: NTT, INTT, and MSM, when the input vector is of size $2^{15}$ and each input bit width is $256$ bits.}
  \vspace{-14pt}
\end{figure}


With the design of this work as the basis, we plan to integrate the module into a system-level application. In the future, we aim to improve elliptic curve computations, both number theoretical transform (NTT) and multi-scalar multiplication (MSM) algorithms, which are essential in the scheme of ZKP. \Cref{fig: ntt-msm-eval} illustrates the scale of memory accesses, modular multiplications, and their intermediate register writes in ZKP components. The values for NTT in \Cref{fig: ntt-msm-eval} are based on simulations of \cite{PNTT_repo}. The values for MSM are calculated using the architecture in \cite{pipezk}. Our work computes large bitwidth modular multiplications efficiently in-SRAM and avoids intermediate register writes and memory accesses, which can significantly improve the performance of ZKP.
\vspace{-6pt}
\section{Conclusion} \label{sec:conclusion}
In this paper, we propose R4CSA-LUT, a new algorithm based on LUTs that combines the merits of both radix-4 modular multiplication and carry save addition in the interleaved algorithm. We also design ModSRAM, an SRAM PIM architecture that aims to compute modular multiplication for ECC based on our co-designed algorithm. The operations in R4CSA-LUT are hardware-friendly and they use LUTs to maximize data reusability. ModSRAM is implemented in state-of-the-art technology and design flow. The experimental results and analysis are based on circuit-level simulation. The area is generated from synthesis and full-custom layout. To the best of our knowledge, we are the first to implement 256-bit modular multiplication in SRAM with 52\% better algorithm efficiency than previous works after scaling with only 32\% area overhead for near-memory computing. We demonstrate a possible solution for combining large-number modular multiplication in SRAM.

\vspace{-8pt}
\begin{acks}
This work is supported by National Science Foundation (NSF) CCF-2328805 and NSF CNS-2112562.
\end{acks}

\bibliographystyle{ACM-Reference-Format}
\bibliography{misc/refs}

\end{document}